\documentclass[apj,twocolumn,iop]{emulateapj}

\usepackage{amsmath}
\usepackage{float}
\usepackage{nicefrac}
\usepackage{rotating}
\usepackage{rotfloat}
\usepackage[none]{hyphenat}
\usepackage{enumitem}
\usepackage{afterpage}
\usepackage{datetime}

\usepackage{array}
\newcolumntype{C}[1]{>{\centering\let\newline\\\arraybackslash\hspace{0pt}}m{#1}}

\usepackage{lineno}
\modulolinenumbers[5]

\usepackage{etoolbox}

\usepackage{listings}
\makeatletter
\long\def\@makecaption#1#2{%
  \vskip\abovecaptionskip
    \footnotesize #1: #2\par
  \vskip\belowcaptionskip}%
\makeatother

\usepackage{color}
\usepackage[flushleft]{threeparttable}

\definecolor{dkgreen}{rgb}{0,0.6,0}
\definecolor{gray}{rgb}{0.5,0.5,0.5}
\definecolor{mauve}{rgb}{0.58,0,0.82}

\usepackage[colorlinks]{hyperref}
\usepackage{etoolbox}

\hypersetup{breaklinks=true,colorlinks=true,linkcolor=blue,citecolor=blue,filecolor=blue,urlcolor=blue}
\makeatletter
\patchcmd{\NAT@citex}
  {\@citea\NAT@hyper@{%
     \NAT@nmfmt{\NAT@nm}%
     \hyper@natlinkbreak{\NAT@aysep\NAT@spacechar}{\@citeb\@extra@b@citeb}%
     \NAT@date}}
  {\@citea\NAT@nmfmt{\NAT@nm}%
   \NAT@aysep\NAT@spacechar\NAT@hyper@{\NAT@date}}{}{}

\patchcmd{\NAT@citex}
  {\@citea\NAT@hyper@{%
     \NAT@nmfmt{\NAT@nm}%
     \hyper@natlinkbreak{\NAT@spacechar\NAT@@open\if*#1*\else#1\NAT@spacechar\fi}%
       {\@citeb\@extra@b@citeb}%
     \NAT@date}}
  {\@citea\NAT@nmfmt{\NAT@nm}%
   \NAT@spacechar\NAT@@open\if*#1*\else#1\NAT@spacechar\fi\NAT@hyper@{\NAT@date}}
  {}{}

\usepackage{array}
\newcolumntype{C}[1]{>{\centering\let\newline\\\arraybackslash\hspace{0pt}}m{#1}}

\def\apj{ApJ}
\def\apjl{ApJ}
\def\apjs{ApJS}
\def\ao{Appl.~Opt.}

\def\aap{A\&A}

\def\mnras{MNRAS}

\def\arcdeg{\hbox{$^\circ$}}

\newcommand{\bc}{\color{blue}}

\makeatletter
\patchcmd{\frontmatter@RRAP@format}{(}{}{}{}
\patchcmd{\frontmatter@RRAP@format}{)}{}{}{}
\renewcommand\Dated@name{}
\makeatother

\newcommand{\altaffilmarkc}[1]{\altaffilmark{\bc #1}}
\newcommand{\emailc}[1]{{\bc #1}}

\shorttitle{\textsc{mpi\_xstar}: MPI-based parallelization of \textsc{xstar}}
\shortauthors{Danehkar et al.}

\begin{document}

\title{MPI\_XSTAR: MPI-based Parallelization of the XSTAR Photoionization Program}

\author{Ashkbiz~Danehkar\altaffilmarkc{1}, 
Michael A. Nowak\altaffilmarkc{2}, 
Julia C. Lee\altaffilmarkc{3}, 
and Randall K. Smith\altaffilmarkc{1}}
\affil{\altaffilmark{1}\,Harvard-Smithsonian Center for Astrophysics,
  60 Garden Street, Cambridge, MA 02138, USA;
  \emailc{ashkbiz.danehkar@cfa.harvard.edu}\\
\altaffilmark{2}\,Massachusetts Institute of Technology, Kavli
Institute for Astrophysics, Cambridge, MA 02139, USA \\
\altaffilmark{3}\,Harvard John A. Paulson School of Engineering \&
Applied Science, 29 Oxford Street, Cambridge, MA 02138 USA 
}

\date[ ]{\footnotesize\textit{Received 2017 October 31; accepted 2017 November 21}}

\begin{abstract}
We describe a program for the parallel implementation of multiple runs of \textsc{xstar}, a photoionization code that is used to predict the physical properties of an ionized gas from its emission and/or absorption lines. The parallelization program, called \textsc{mpi\_xstar}, has been developed and implemented in the C++ language by using the Message Passing Interface (MPI) protocol, a conventional standard of parallel computing. We have benchmarked parallel multiprocessing executions of \textsc{xstar}, using \textsc{mpi\_xstar}, against a serial execution of \textsc{xstar}, in terms of the parallelization speedup and the computing resource efficiency. Our experience indicates that the parallel execution runs significantly faster than the serial execution, however, the efficiency in terms of the computing resource usage decreases with increasing the number of processors used in the parallel computing.
\end{abstract}


\section{Introduction}
\label{mpixstar:introduction}

The Fortran 77 code \textsc{xstar} \citep{Kallman1996,Kallman2001,Kallman2004,Kallman2009} 
is a photoionization program used to model the X-ray spectra from astrophysical plasmas. \textsc{xstar} was developed to calculate ionization structures, thermal structures, and emissivities of a spherical gaseous cloud with given physical conditions and elemental abundances being ionized by a central source of ionizing radiation that is specified by its luminosity and spectral continuum. It uses 
an atomic database \citep{Bautista2001} that includes energy levels, recombination rate coefficients, transition probabilities, and cross sections of most of the lines seen in the UV and X-ray bands.
To facilitate model fitting for various physical parameters, a Perl script, called \textsc{xstar2xspec}, was developed and is included in \textsc{xstar} that executes multiple serial runs of \textsc{xstar}, and utilizes the results of these \textsc{xstar} computations to generate multiplicative tabulated models for spectroscopic analysis tools. These tabulated models are the  absorption spectrum  table (\textsf{xout\_}\textsf{mtable.fits}), the reflected emission spectrum table (\textsf{xout\_}\textsf{ain.fits}), and the transmitted emission spectrum table in the absorption direction (\textsf{xout\_}\textsf{aout.fits}), which are used for the photoionization model fitting in X-ray spectral modeling tools such as \textsc{xspec} \citep{Arnaud1996} and \textsc{isis} \citep{Houck2000}.

\textsc{xstar} is widely used to obtain physical conditions of X-ray ionized emitters and/or absorbers in active galactic nuclei \citep[AGNs; e.g.][]{Brenneman2011,Tombesi2011a,Tombesi2011b,Tombesi2014,Lee2013,Laha2014} and non-equilibrium ionized gaseous clouds surrounding X-ray binaries \citep[XRBs; e.g.][]{Neilsen2009a,Neilsen2011,Miller2012,Miller2015}. Fundamental physical parameters in typical models are the gas number density $n$, the hydrogen column density $N_{\rm H}= n_{\rm H} V_{f} \triangle r$, and the ionization parameter $\xi=L_{\rm ion}/n_{\rm H} r^2$ \citep{Tarter1969}, where $L_{\rm ion}$ is the luminosity (1--1000 Ryd) of the ionizing source, $V_{f}$ is the volume filling factor of the gaseous cloud, $r$ is the distance from the central source of ionizing radiation and $\triangle r$  is the thickness of the gaseous cloud. In most cases, the gas number density and chemical abundances are fixed, while the variations of the hydrogen column density ($N_{\rm H}$) and the ionization parameter ($\xi$) are used to constrain the physical properties of an ionized gas. Hence, multiplicative tabulated models are generated on the two-dimensional $N_{\rm H}$--$\xi$ plane, sampling the column density with $n$ intervals and the ionization parameter with $m$ intervals, which require $n \times m$ times \textsc{xstar} runs. Taking our typical one-hour \textsc{xstar} run (for our specified model settings), \textsc{xstar2xspec} takes 600 hours (25 days) for $n=20$ and $m=30$ to produce tabulated models.

To facilitate parallel executions of multiple \textsc{xstar} runs, a Unix shell script together with some S-Lang codes, so-called \textsc{pvm\_xstar}\footnote{\url{http://space.mit.edu/cxc/pvm_xstar/}} \citep{Noble2009}, were developed based on the Parallel Virtual Machine (PVM) library\footnote{\url{http://www.csm.ornl.gov/pvm/}} \citep{Geist1994} and the S-Lang PVM module\footnote{\url{http://space.mit.edu/cxc/modules/pvm/}} \citep{Davis2005,Noble2006}. \textsc{pvm\_xstar} loads the S-Lang PVM module, spawns multiple slave processes for \textsc{xstar} runs using the job list created by \textsc{xstinitable}, and invokes \textsc{xstar2table} on the results of these \textsc{xstar} runs to create table models. The program \textsc{xstinitable} is a code written in the C language, and is included in the \textsc{ftools} package\footnote{\url{https://heasarc.gsfc.nasa.gov/ftools/}} which is used in the initialization step of the Perl script \textsc{xstar2xspec} to generate the job list file (xstinitable.lis), containing \textsc{xstar} calling commands for the variation of the physical parameters used in photoionization modeling, and to create an initial FITS file (\textsf{xstinitable.fits}) required for producing multiplicative tabulated models. The program \textsc{xstar2table} is also written in the C language, and included in the \textsc{ftools} package which is invoked by the Perl script \textsc{xstar2xspec} to generate tabulated models from multiple serial runs of \textsc{xstar}. \textsc{pvm\_xstar}, similar to \textsc{xstar2xspec}, invokes \textsc{xstinitable} and \textsc{xstar2table} from the \textsc{ftools} package, but executes those \textsc{xstar} calling commands in parallel. However, the PVM software package is not employed by modern supercomputers. The two most common protocols of parallel computing used by recent supercomputers are Message Passing Interface (MPI) and OpenMP. The MPI protocol is for clusters with distributed memory, while OpenMP supports shared memory systems. Since it may not be possible to use \textsc{pvm\_xstar} on a given computer cluster, it was necessary to develop a code that permits parallel executions of multiple \textsc{xstar} runs using either MPI or OpenMP. As the MPI protocol is used by all modern clusters with distributed memory, we attempted to implement an MPI-based interface for parallelizing the \textsc{xstar} program.

This paper presents a parallelization implementation for multiple runs of the \textsc{xstar} program, so called \textsc{mpi\_xstar}, developed using the MPI library in the C++ language. 
Section~\ref{mpixstar:method} describes the design of the mpi\_xstar program. Section~\ref{mpixstar:result} evaluates the parallelization speedup and the computing resource efficiency of mpi\_xstar, followed by a conclusion in Section~\ref{mpixstar:conclusion}.

\section{Parallel MPI Implementation}
\label{mpixstar:method}

\textsc{mpi\_xstar}, a parallel-manager program, has been written in C++ using the high-performance computing industry standard MPI library \citep[e.g..][]{Gropp1999}. The \textsc{mpi\_xstar} source code is freely available under GPLv3 license on GitHub\footnote{\url{https://github.com/xstarkit/MPI_XSTAR}}.  \textsc{mpi\_xstar} is designed to be used on a cluster or a  multi-core machine composed of multiple CPUs (central processor units) that enables an easy generation of multiplicative tabulated models for spectral modeling tools. The code has initially been developed and examined on the \textsc{odyssey} cluster at Harvard University. \textsc{mpi\_xstar} utilizes \textsc{xstinitable} and \textsc{xstar2table}, similar to \textsc{xstar2xspec} and \textsc{pvm\_xstar}. It uses the MPI library to run the \textsc{xstar} calling commands on a number of CPUs (e.g., specified by the \textsf{np} parameter value of \textsf{mpirun}). 

An outline of the implementation of an MPI-based parallelization of \textsc{xstar}, \textsc{mpi\_xstar}, is
summarized in a pseudo-code in Algorithm~\ref{algorithm1}. 
The program begins with the generation of \textsc{xstar} calling commands for the variation of the input parameters and an initial FITS file using \textsc{xstinitable} in the master processor (rank zero).  It passes the command argument [returned by an {\small\tt{MPI\_Init()}}  call] to \textsc{xstinitable} to create the \textsc{xstar} calling commands file (xstinitable.lis) and the initial FITS file (xstinitable.fits). It makes three copies of the initial FITS file, which will be used by \textsc{xstar2table} at the end to create three table files (\textsf{xout\_}\textsf{mtable.fits}, \textsf{xout\_}\textsf{ain.fits}, and \textsf{xout\_}\textsf{aout.fits}). While the master processor performs the initialization for a parallel setup of \textsc{xstar}, other processors are blocked [by using an {\small\tt{MPI\_Barrier()}}  call] and waited. After the master processor completes its initialization, it broadcasts the number of \textsc{xstar} calling commands  to other processors [by an {\small\tt{MPI\_Bcast()}}  call], that allows each processor to know which \textsc{xstar} calling command should be executed.

\begin{lstlisting}[caption={Pseudo-code structure of the \textsc{mpi\_xstar} program}, label=algorithm1,language=VBScript]
call MPI_Init(argument)
call MPI_Comm_rank(processor_rank)
call MPI_Comm_size(total_processor_number)

if proc_rank = 0 then 
	call xstinitable to initialize the job list 

call MPI_Barrier()

call MPI_Bcast(total_job_number)

job_number = processor_rank + 1

loop job_number <= total_job_number
	read the xstar command from the job list 
	call xstar on the command
	job_number = job_number + 1
	call MPI_Bcast(job_number)
	
call MPI_Barrier()

if proc_rank = 0 then 
	call xstar2table to make the FITS table models 
	
call MPI_Barrier()
call MPI_Finalize()
\end{lstlisting}

After each processor is allocated to an initial calling order based on the
rank value [returned by an {\small\tt{MPI\_Comm\_rank()}} call], the program begins to read the \textsc{xstar} calling command from the initialization file (xstinitable.lis) and execute it by calling the \textsc{xstar} program. The calling order of \textsc{xstar} commands, which is being executed by a processor, is sent to other processors [by an {\small\tt{MPI\_Bcast()}} call], so it will not be run by them. Each processor is being blocked until its \textsc{xstar} computation is done. After a processor completes its current \textsc{xstar} command, it then reads the next calling command and executes a new \textsc{xstar} command. When there is no calling command for execution by a processor, those processors, which finish their task, are blocked [by an {\small\tt{MPI\_Barrier()}}  call] and waited for those processors whose  \textsc{xstar} commands are still running. 

After all \textsc{xstar} calling commands are executed, the master processor invokes \textsc{xstar2table} upon the results of these \textsc{xstar} computations in order to generate multiplicative tabulated models, while other processors are blocked [by an {\small\tt{MPI\_Barrier()}}  call]. The \textsc{ftools} program \textsc{xstar2table} uses the \textsc{xstar} outputs in each folder, and adds them to table files (\textsf{xout\_}\textsf{mtable.fits}, \textsf{xout\_}\textsf{ain.fits}, and \textsf{xout\_}\textsf{aout.fits}) created from the initial FITS file in the initialization step. For a job list containing $n$ \textsc{xstar} calling commands, it is required to execute $n$ runs of \textsc{xstar2table} upon the results in each folder to generate multiplicative tabulated models. As this procedure is very quick, it is done by the master processor in a serial mode (on a single CPU). After tabulated model FITS files are produced by the master processor, all processors are unblocked and are terminated [by an {\small\tt{MPI\_Finalize()}}  call].  
\textsc{mpi\_xstar} outputs include the standard output log file (\textsc{xstar2xspec.log}; similar to the script \textsc{xstar2xspec}) and the error report file for certain failure conditions such as the absence of the xstar commands file (xstinitable.lis) and the initial FITS file (xstinitable.fits).

\section{Computing Performance}
\label{mpixstar:result}

To evaluate the \textsc{mpi\_xstar} computing performance, we calculated a grid of $9 \times 6$ \textsc{xstar} models for the two-dimensional $N_{\rm H}$--$\xi$ parameter space, sampling the column density with 9 logarithmic intervals and an interval size of $0.5$ from $\log N_{\rm H}=20$ to $24$\,cm$^{-2}$, and the ionization parameter  with 6 logarithmic intervals and an interval size of $1$ from $\log\xi=0$ to $5$ erg\,cm\,s$^{-1}$, assuming a gas density of $\log n=12$\,cm$^{-3}$ and a turbulent velocity of $v_{\rm turb}=100$\,km\,s$^{-1}$. We assumed a spherical geometry with a covering fraction of $C_{f} =\Omega / 4 \pi= 0.4$. The chemical composition is assumed to be solar elemental abundances \citep[$A_{\rm Fe} = 1$;][]{Grevesse1996}. The initial gas temperature of $T_{\rm init}=10^6$\,K used here is typical for AGNs \citep[][]{Nicastro1999,Bianchi2005}. The parameters used for \textsc{mpi\_xstar} benchmarks are listed in Table~\ref{pg1211:fit:input}. We also employed a spectral energy distribution (SED) described in \citet{Danehkar2017} as the central source of ionizing radiation with a typical luminosity of $L_{\rm ion}=10^{44}$~erg\,s$^{-1}$ (between 1 and 1000 Ryd). We run \textsc{mpi\_xstar} on the Harvard \textsc{odyssey} cluster, consisting of 60,000 cores and 4 GB of memory per core on average, running the CentOS v6.5 implementation of the Linux operating system and scheduling the jobs using the SLURM v16.05 resource manager.  To calculate the speedup and efficiency, we have submitted different \textsc{mpi\_xstar} jobs with a single CPU and multiple CPUs (2 to 54).

\begin{table}
\footnotesize
\caption{Parameters used for \textsc{mpi\_xstar} benchmark model.
\label{pg1211:fit:input}
}
\begin{center}
\begin{tabular}{C{4.2cm}cc}
\hline\hline
\noalign{\smallskip}
{Parameter} & {Value}  & {Interval Size}  \\
\noalign{\smallskip}
\hline 
\noalign{\smallskip}
$\log N_{\rm H}$ (cm$^{-3}$)\dotfill  & $20 \cdots 24$   & $0.5$\\ 
\noalign{\smallskip}
$\log \xi$ (erg\,cm\,s$^{-1}$)\dotfill  & $0 \cdots 5$  & $1.0$ \\
\noalign{\smallskip}
$\log n$ (cm$^{-3}$)\dotfill  & $12$   & --\\ 
\noalign{\smallskip}
$v_{\rm turb}$ (km\,s$^{-1}$)\dotfill  & $100$  & -- \\ 
\noalign{\smallskip}
$C_{f}=\Omega / 4 \pi$\dotfill  & $ 0.4$   & --\\ 
\noalign{\smallskip}
$A_{\rm Fe}$ \dotfill  & $ 1.0$   & --\\ 
\noalign{\smallskip}
$T_{\rm init}$ ($10^{4}$ K)\dotfill  & $ 100$   &  --\\ 
\noalign{\smallskip}
$L_{\rm ion}$ ($10^{38}$ erg\,s$^{-1}$)\dotfill  & $1.0\times 10^{6}$  & --\\ 
\noalign{\smallskip} 
\hline
\noalign{\smallskip}
\end{tabular}
\end{center}
\begin{tablenotes}
\footnotesize
\item[1]\textbf{Notes.} Logarithmic interval sizes are chosen for the total hydrogen column density ($N_{\rm H}$) and the ionization parameter ($\xi=L_{\rm ion}/n_{\rm H} r^2$).
\end{tablenotes}
\end{table}

The speedup $\mathcal{S}(N)$ of a parallel computation with $N$ processors is defined as follows:
\begin{equation}
\mathcal{S}(N) \equiv \frac{\mathcal{T}(1)}{\mathcal{T}(N)},
\label{eq_1}
\end{equation}
where $\mathcal{T}(i)$ is the running time for a parallel execution with $i$ processors, so $\mathcal{T}(1)$ corresponds to a serial execution. The speedup for a single processor ($N = 1$) is defined to be $\mathcal{S}(1)=1$. Ideally, an excellent speedup is achieved when $\mathcal{S}(N)\approx N$. 

The efficiency $\mathcal{E}(N)$ in using the computing resources for a parallel computation with $N$ processors is defined as follows:
\begin{equation}
\mathcal{E}(N) \equiv \frac{\mathcal{S}(N)}{N} = \frac{\mathcal{T}(1)}{N \times \mathcal{T}(N)}.
\label{eq_2}
\end{equation}
The efficiency is typically between zero and one. An efficiency of more than one describes the so-called superlinear speedup. As the speedup $\mathcal{S}(1)=1$, the efficiency is $\mathcal{E}(1)=1$ for a single processor ($N=1$).

Table~\ref{mpixstar:speedup} lists the running time, the speedup, the efficiency of \textsc{mpi\_xstar} with 1 to 54 CPUs. It can been seen that the running time $\mathcal{T}(N)$ of the parallel executions is significantly shorter than the
serial execution. It took around 18 hours to make \textsc{xstar} grid models with 32 and 54 CPUs, whereas about 10 days using a single CPU ($N=1$). Although the speedup $\mathcal{S}(N)$ increases with the number of processors ($N$), it does not demonstrate an ideal speedup ($\mathcal{S}(N)\approx N$). We also notice that the efficiency $\mathcal{E}(N)$ decreases with increasing the number of processors ($N$). 

\begin{figure}
\begin{center}
\includegraphics[width=3.in, trim = 45 30 0 0, clip, angle=0]{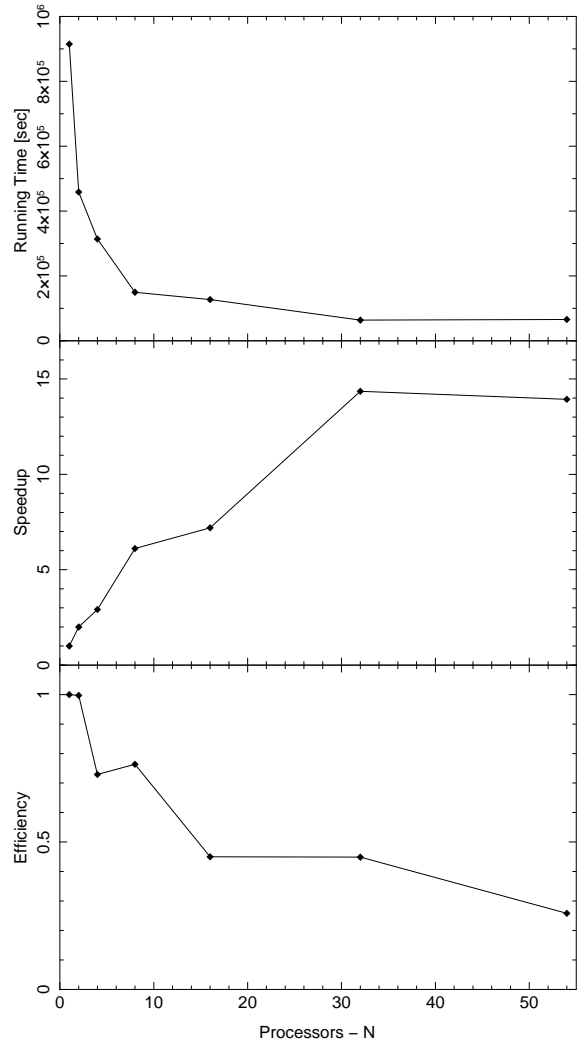}%
\caption{From top to bottom, the running time $\mathcal{T}(N)$, the speedup $\mathcal{S}(N)$, the efficiency $\mathcal{E}(N)$ and as a function of the number of processes ($N$) for a benchmark photoionization model with the parameters listed in Table~\ref{pg1211:fit:input}. The running time $\mathcal{T}(N)$ is in seconds. 
\label{fig:performance}%
}
\end{center}
\end{figure}

\begin{table}
\footnotesize
\caption{Running time, speedup and efficiency of \textsc{mpi\_xstar}
\label{mpixstar:speedup}
}
\begin{center}
\begin{tabular}{cccc}
\hline\hline
\noalign{\smallskip}
{~$N$~} & {~$\mathcal{T}(N)$~} & {~$\mathcal{S}(N)$~}  & {~$\mathcal{E}(N)$~}  \\
\noalign{\smallskip}
\hline 
\noalign{\smallskip}
1    &   254:07:25  &  1.00   & 1.00    \\ 
\noalign{\smallskip}
2    &  127:23:58   &  1.99   & 0.99    \\ 
\noalign{\smallskip}
4    &  87:09:47   &  2.92   &  0.73   \\ 
\noalign{\smallskip}
8    &  41:35:41   &  6.11   &  0.76   \\ 
\noalign{\smallskip}
16   & 35:18:21    &  7.20   &  0.45   \\ 
\noalign{\smallskip}
32   & 17:42:15    &  14.35   &  0.45   \\ 
\noalign{\smallskip}
54   & 18:13:30    &  13.93   &  0.26   \\ 
\noalign{\smallskip}
\hline
\noalign{\smallskip}
\end{tabular}
\end{center}
\begin{tablenotes}
\footnotesize
\item[1]\textbf{Notes.} The running time $\mathcal{T}(N)$ is in hours, minutes, and seconds (hh:mm:ss).
\end{tablenotes}
\end{table}

The performance results for \textsc{mpi\_xstar} versus $N$ processors are shown in Fig.~\ref{fig:performance}, including the running time $\mathcal{T}(N)$, the speedup $\mathcal{S}(N)$ and the efficiency $\mathcal{E}(N)$. As seen in the figure, the speedup and efficiency are not linearly correlated with the number of processors. This is due to the fact that the running time of each \textsc{xstar} process greatly varies according to the physical conditions ($\log N_{\rm H}$ and $\log \xi$), so they are not identical to each other. We notice that the running time of a parallel execution is limited by the maximum running time of the \textsc{xstar} program for given physical parameters. For our benchmark example, it took between 25 seconds and 17.5 hours for each \textsc{xstar} run, depending on the column density $\log N_{\rm H}$ and the ionization parameter $\log\xi$  used as input parameters. Parallel running times of multiple \textsc{xstar} runs do not exceed the maximum running time of a single \textsc{xstar}. There should not be much difference between the parallel executions with $N=32$ and $54$.
However, as seen in Table~\ref{mpixstar:speedup}, the parallel computing with $N=54$ is roughly a half hour longer than that with $N=32$. This is due to the fact that each cluster node (of the Harvard \textsc{odyssey}) used in our benchmark consists of 32 cores, so we had to use 2 nodes for the parallel computing with $N=54$. The inter-node communication time slightly makes two-node parallel-computing (more than 32 CPUs) slower than single-node parallel-computing (only with 32 CPUs).

As the execution time of each single \textsc{xstar} restricts the parallel running time of \textsc{mpi\_xstar}, it prevents us from achieving a prefect speedup ($\mathcal{S}(N)\approx N$). If the internal Fortran 77 routines of the program \textsc{xstar} were extended according to one of convention protocols of parallel computing (MPI or OpenMP), an ideal speedup might be achievable. Nevertheless, despite the low computing efficiency of \textsc{mpi\_xstar}, it provides a major improvement for constructing  photoionization grid models for spectroscopic fitting tools such as \textsc{xspec} and \textsc{isis}. For example, the photoionization table model with the settings listed in Table~\ref{pg1211:fit:input} can be now produced in 18 hours using a parallel execution with 32 CPUs rather than 10 days using a serial execution.

\section{Conclusion}
\label{mpixstar:conclusion}

This paper presents \textsc{mpi\_xstar}, which is a parallel implementation of multiple \textsc{xstar} runs using the MPI protocol \citep[e.g..][]{Gropp1999} for clusters with distributed memory. \textsc{mpi\_xstar} expedites the tabulated model generation on the high-performance computing environments. \textsc{mpi\_xstar}, similar to \textsc{xstar2xspec} and \textsc{pvm\_xstar}, invokes the \textsc{ftools} programs \textsc{xstinitable} and \textsc{xstar2table}. The table models take an extremely long time to be produced by \textsc{xstar2xspec}. Moreover, \textsc{pvm\_xstar} relies on on the PVM technology \citep{Geist1994}, which is no longer supported by modern supercomputers. Hence, an MPI-based manager for parallelizing \textsc{xstar} can overcome the current difficulties in producing the multiplicative tabulated models. 

The \textsc{mpi\_xstar} code that we have developed is available via GitHub (\url{github.com/xstarkit/mpi_xstar}).  Note that it makes use of the locally installed \textsc{xstar} and its associated tools, and will run regardless of \textsc{xstar} version so long as the \textsc{xstar}  parameter inputs and calling sequence do not change.  However, should newer versions of \textsc{xstar} arise requiring such changes, updates to the \textsc{mpi\_xstar} code will be made and documented on the GitHub site.

The code was evaluated for the generation of the \textsc{xstar} table models with a grid of $9 \times 6$ on the $N_{\rm H}$--$\xi$ parameter space. The parallel multiprocessing execution is significantly faster than the serial execution, as the computation,  which previously took 10 days, requires only about 18 hours using 32 CPUs. However, our benchmarking studies with 1 to 54 CPUs indicates that the parallel efficiency decreases with increasing the number of processors. Moreover, we did not find any linear correlation between the speedup and the number of processors, as shown in Fig.~\ref{fig:performance}.  
Although we did not achieved an ideal speedup ($\mathcal{S}(N)\approx N$), the running times (see Table~\ref{mpixstar:speedup}) of parallel execution with 32 and 54 CPUs are enormously shorter than the time of a serial execution. We notice that the performance of \textsc{mpi\_xstar} is restricted by the maximum running time of a single \textsc{xstar} run (about 17.5 hours for our benchmark model results listed in Table~\ref{mpixstar:speedup}). However, \textsc{mpi\_xstar} provides a faster way for the generation of photoionization grid models for spectral model fitting tools such as \textsc{xspec} \citep{Arnaud1996} and \textsc{isis} \citep{Houck2000}. 

In summary, the new code \textsc{mpi\_xstar} is able to speed up the photoionization modeling procedure. An important application is a fast generation of photoionization table models of X-ray warm absorbers in AGNs \citep[e.g.,][]{Danehkar2017}, whose computation, depending on the number of CPUs requested for the parallel execution, is shorter than a serial execution using \textsc{xstar2xspec}. The parallelization of \textsc{xstar} might be implementable on the graphical processing units (GPU) using
the CUDA library. Moreover. it might be possible to parallelize the internal routines (currently in Fortran 77) of the program \textsc{xstar}, which will significantly expedite photoionization simulations of ionized gaseous clouds. An MPI-CUDA GPU-based parallelization and rewriting the \textsc{xstar} internal routines based on the MPI library for the high-performance computing environments deserve further investigations  in the future.

\acknowledgments

This work was supported by NASA through the CXO grant GO5-16108X. 
We thank the referee, Tim Kallman, for his careful reading and valuable suggestions. 
The computations in this paper were run on the Odyssey cluster supported by the FAS Division of Science, Research Computing Group at Harvard University, the Smithsonian Institution High Performance Cluster (Hydra), and the Stampede cluster at the Texas Advanced Computing Center supported by NSF grant ACI-1134872.

%\bibliographystyle{apj}                       %% AASTeX
%\bibliography{references}

\begin{thebibliography}{27}
\expandafter\ifx\csname natexlab\endcsname\relax\def\natexlab#1{#1}\fi

\bibitem[{{Arnaud}(1996)}]{Arnaud1996}
{Arnaud}, K.~A. 1996, in Astronomical Society of the Pacific Conference Series,
  Vol. 101, Astronomical Data Analysis Software and Systems V, ed. G.~H.
  {Jacoby} \& J.~{Barnes}, 17

\bibitem[{{Bautista} \& {Kallman}(2001)}]{Bautista2001}
{Bautista}, M.~A. \& {Kallman}, T.~R. 2001, \apjs, 134, 139

\bibitem[{{Bianchi} {et~al.}(2005){Bianchi}, {Matt}, {Nicastro}, {Porquet}, \&
  {Dubau}}]{Bianchi2005}
{Bianchi}, S., {Matt}, G., {Nicastro}, F., {et~al.} 2005, \mnras, 357, 599

\bibitem[{{Brenneman} {et~al.}(2011){Brenneman}, {Reynolds}, {Nowak}, {Reis},
  {Trippe}, {Fabian}, {Iwasawa}, {Lee}, {Miller}, {Mushotzky}, {Nandra}, \&
  {Volonteri}}]{Brenneman2011}
{Brenneman}, L.~W., {Reynolds}, C.~S., {Nowak}, M.~A., {et~al.} 2011, \apj,
  736, 103

\bibitem[{{Danehkar} {et~al.}(2017){Danehkar}, {Nowak}, {Lee}, {Kriss},
  {Young}, {Hardcastle}, {Chakravorty}, {Fang}, {Neilsen}, {Rahoui}, \&
  {Smith}}]{Danehkar2017}
{Danehkar}, A., {Nowak}, M.~A., {Lee}, J.~C., {et~al.} 2017, \apj, submitted

\bibitem[{{Davis} {et~al.}(2005){Davis}, {Houck}, {Allen}, \&
  {Stage}}]{Davis2005}
{Davis}, J.~E., {Houck}, J.~C., {Allen}, G.~E., {et~al.} 2005, in Astronomical
  Society of the Pacific Conference Series, Vol. 347, Astronomical Data
  Analysis Software and Systems XIV, ed. P.~{Shopbell}, M.~{Britton}, \&
  R.~{Ebert}, 444

\bibitem[{{Geist} {et~al.}(1994){Geist}, {Beguelin}, {Dongarra}, {Jiang},
  {Manchek}, \& {Sunderam}}]{Geist1994}
{Geist}, A., {Beguelin}, A., {Dongarra}, J., {et~al.} 1994, PVM: Parallel
  Virtual Machine, A Users' Guide and Tutorial for Networked Parallel
  Computing, ed. J.~{Kowalik} (MIT Press, Scientific and Engineering
  Computation)

\bibitem[{{Grevesse} {et~al.}(1996){Grevesse}, {Noels}, \&
  {Sauval}}]{Grevesse1996}
{Grevesse}, N., {Noels}, A., \& {Sauval}, A.~J. 1996, in Astronomical Society
  of the Pacific Conference Series, Vol.~99, Cosmic Abundances, ed. S.~S.
  {Holt} \& G.~{Sonneborn}, 117

\bibitem[{{Gropp} {et~al.}(1999){Gropp}, {Lusk}, \& {Skjellum}}]{Gropp1999}
{Gropp}, W., {Lusk}, E., \& {Skjellum}, A. 1999, {Using MPI: Portable
  Programming with the Message-Passing Interface}

\bibitem[{{Houck} \& {Denicola}(2000)}]{Houck2000}
{Houck}, J.~C. \& {Denicola}, L.~A. 2000, in Astronomical Society of the
  Pacific Conference Series, Vol. 216, Astronomical Data Analysis Software and
  Systems IX, ed. N.~{Manset}, C.~{Veillet}, \& D.~{Crabtree}, 591

\bibitem[{{Kallman} \& {Bautista}(2001)}]{Kallman2001}
{Kallman}, T. \& {Bautista}, M. 2001, \apjs, 133, 221

\bibitem[{{Kallman} {et~al.}(2009){Kallman}, {Bautista}, {Goriely}, {Mendoza},
  {Miller}, {Palmeri}, {Quinet}, \& {Raymond}}]{Kallman2009}
{Kallman}, T.~R., {Bautista}, M.~A., {Goriely}, S., {et~al.} 2009, \apj, 701,
  865

\bibitem[{{Kallman} {et~al.}(1996){Kallman}, {Liedahl}, {Osterheld},
  {Goldstein}, \& {Kahn}}]{Kallman1996}
{Kallman}, T.~R., {Liedahl}, D., {Osterheld}, A., {et~al.} 1996, \apj, 465, 994

\bibitem[{{Kallman} {et~al.}(2004){Kallman}, {Palmeri}, {Bautista}, {Mendoza},
  \& {Krolik}}]{Kallman2004}
{Kallman}, T.~R., {Palmeri}, P., {Bautista}, M.~A., {et~al.} 2004, \apjs, 155,
  675

\bibitem[{{Laha} {et~al.}(2014){Laha}, {Guainazzi}, {Dewangan}, {Chakravorty},
  \& {Kembhavi}}]{Laha2014}
{Laha}, S., {Guainazzi}, M., {Dewangan}, G.~C., {et~al.} 2014, \mnras, 441,
  2613

\bibitem[{{Lee} {et~al.}(2013){Lee}, {Kriss}, {Chakravorty}, {Rahoui}, {Young},
  {Brandt}, {Hines}, {Ogle}, \& {Reynolds}}]{Lee2013}
{Lee}, J.~C., {Kriss}, G.~A., {Chakravorty}, S., {et~al.} 2013, \mnras, 430,
  2650

\bibitem[{{Miller} {et~al.}(2015){Miller}, {Fabian}, {Kaastra}, {Kallman},
  {King}, {Proga}, {Raymond}, \& {Reynolds}}]{Miller2015}
{Miller}, J.~M., {Fabian}, A.~C., {Kaastra}, J., {et~al.} 2015, \apj, 814, 87

\bibitem[{{Miller} {et~al.}(2012){Miller}, {Pooley}, {Fabian}, {Nowak}, {Reis},
  {Cackett}, {Pottschmidt}, \& {Wilms}}]{Miller2012}
{Miller}, J.~M., {Pooley}, G.~G., {Fabian}, A.~C., {et~al.} 2012, \apj, 757, 11

\bibitem[{{Neilsen} {et~al.}(2009){Neilsen}, {Lee}, {Nowak}, {Dennerl}, \&
  {Vrtilek}}]{Neilsen2009a}
{Neilsen}, J., {Lee}, J.~C., {Nowak}, M.~A., {et~al.} 2009, \apj, 696, 182

\bibitem[{{Neilsen} {et~al.}(2011){Neilsen}, {Remillard}, \&
  {Lee}}]{Neilsen2011}
{Neilsen}, J., {Remillard}, R.~A., \& {Lee}, J.~C. 2011, \apj, 737, 69

\bibitem[{{Nicastro} {et~al.}(1999){Nicastro}, {Fiore}, \&
  {Matt}}]{Nicastro1999}
{Nicastro}, F., {Fiore}, F., \& {Matt}, G. 1999, \apj, 517, 108

\bibitem[{{Noble} {et~al.}(2006){Noble}, {Houck}, {Davis}, {Young}, \&
  {Nowak}}]{Noble2006}
{Noble}, M.~S., {Houck}, J.~C., {Davis}, J.~E., {et~al.} 2006, in Astronomical
  Society of the Pacific Conference Series, Vol. 351, Astronomical Data
  Analysis Software and Systems XV, ed. C.~{Gabriel}, C.~{Arviset}, D.~{Ponz},
  \& S.~{Enrique}, 481

\bibitem[{{Noble} {et~al.}(2009){Noble}, {Ji}, {Young}, \& {Lee}}]{Noble2009}
{Noble}, M.~S., {Ji}, L., {Young}, A., {et~al.} 2009, in Astronomical Society
  of the Pacific Conference Series, Vol. 411, Astronomical Data Analysis
  Software and Systems XVIII, ed. D.~A. {Bohlender}, D.~{Durand}, \&
  P.~{Dowler}, 301

\bibitem[{{Tarter} {et~al.}(1969){Tarter}, {Tucker}, \&
  {Salpeter}}]{Tarter1969}
{Tarter}, C.~B., {Tucker}, W.~H., \& {Salpeter}, E.~E. 1969, \apj, 156, 943

\bibitem[{{Tombesi} {et~al.}(2011{\natexlab{a}}){Tombesi}, {Cappi}, {Reeves},
  {Palumbo}, {Braito}, \& {Dadina}}]{Tombesi2011a}
{Tombesi}, F., {Cappi}, M., {Reeves}, J.~N., {et~al.} 2011{\natexlab{a}}, \apj,
  742, 44

\bibitem[{{Tombesi} {et~al.}(2011{\natexlab{b}}){Tombesi}, {Sambruna},
  {Reeves}, {Reynolds}, \& {Braito}}]{Tombesi2011b}
{Tombesi}, F., {Sambruna}, R.~M., {Reeves}, J.~N., {et~al.} 2011{\natexlab{b}},
  \mnras, 418, L89

\bibitem[{{Tombesi} {et~al.}(2014){Tombesi}, {Tazaki}, {Mushotzky}, {Ueda},
  {Cappi}, {Gofford}, {Reeves}, \& {Guainazzi}}]{Tombesi2014}
{Tombesi}, F., {Tazaki}, F., {Mushotzky}, R.~F., {et~al.} 2014, \mnras, 443,
  2154

\end{thebibliography}

\end{document}